\documentclass[a4paper,11pt]{article}
\usepackage{pos}

\usepackage{amsmath, bbm}                      
\usepackage{graphicx}
\usepackage{float}
\usepackage{svg}
\usepackage{subcaption}
\usepackage{url}
\usepackage{tikz}
\usepackage{transparent}
\usepackage[compat=1.1.0]{tikz-feynman}
\usepackage{wrapfig}
\def\verbatim@font{\linespread{1}\normalfont\ttfamily}
\makeatother

\usepackage{xspace}

\newcommand{\pythia}{\textsc{Pythia}\xspace}
\newcommand{\vincia}{\textsc{Vincia}\xspace}
\newcommand{\sherpa}{\textsc{Sherpa}\xspace}

\newcommand{\rivet}{\textsc{Rivet}\xspace}

\title{Multi-Jet Merging in Deep Inelastic Scattering with \pythia}

\author*[a,b]{Joni Laulainen}
\author[a,b]{Ilkka Helenius}
\author[c]{Christian T. Preuss}

\affiliation[a]{University of Jyvaskyla, Department of Physics, \\ P.O. Box 35, FI-40014 University of Jyvaskyla, Finland}

\affiliation[b]{Helsinki Institute of Physics, \\ P.O. Box 64, FI-00014 University of Helsinki, Finland}

\affiliation[c]{Department of Physics, University of Wuppertal, 42119 Wuppertal, Germany}

\emailAdd{joollaul@jyu.fi}
\emailAdd{ilkka.m.helenius@jyu.fi}
\emailAdd{preuss@uni-wuppertal.de}

\abstract{We aim to improve the modelling of deep inelastic scattering by implementing multi-jet merging capabilities in Vincia parton shower in the general-purpose Monte Carlo event generator, \pythia. Merging allows to combine event samples of different parton multiplicities with logarithmically enhanced radiation from parton shower algorithms, without double-counting. Here we consider events up to five outgoing partons and present results for jet analyses compared to experimental data provided by the ZEUS and H1 collaborations of the HERA collider. The analyses span a wide range in photon virtuality, and the results show that the multi-jet merging improves jet modelling especially for low virtuality events.
}

\FullConference{31st International Workshop on Deep Inelastic Scattering (DIS2024)\\
 8–12 April 2024\\
Grenoble, France\\}

\begin{document}
\maketitle

\section{Introduction}

\begin{figure}[h]
\centering
    \begin{tikzpicture}[opacity=1.0,thick,scale=0.7, every node/.style={transform shape}]
        \begin{feynman}[large]
            \vertex (a) {\(k\)};
            \vertex [right=of a] (b);
            \vertex [above right=of b] (f1) {\(k'\)};
            \vertex [below right=of b] (c);
            \vertex [right=of c] (f2) {\(\hat{p}_1\)};
            \vertex [below=of c] (d);
            \vertex [below left=of d] (e);
            \vertex [right=of d] (f3) {\(\hat{p}_2\)};
            \vertex [left=of e] (f) {\(p\)};;
            \vertex (w1) at (6, -4.4);
            \vertex (w2) at (6, -4.83) {\(X\)};
            \vertex (w3) at (6, -5.2);
            
            \node [dot, scale=2.5] (node) at (2.1,-4.83);
            
            \diagram* {
                (a) -- [fermion] (b) -- [fermion] (f1),
                (b) -- [boson, edge label=\(q\)]  (c),
                (c) -- [fermion] (f2),
                (f) -- [fermion] (e) [blob] -- [fermion, edge label=\(\hat{p}\)] (d),
                (d) -- [fermion] (c),
                (d) -- [gluon] (f3),
                (e) -- [fermion] (w1),
                (e) -- [fermion] (w2),
                (e) -- [fermion] (w3),
            };
        \end{feynman}
    \end{tikzpicture}
    \caption{An example of a Feynman diagram of a dijet event of DIS process.}
    \label{fig001:DISFeynman}
\end{figure}
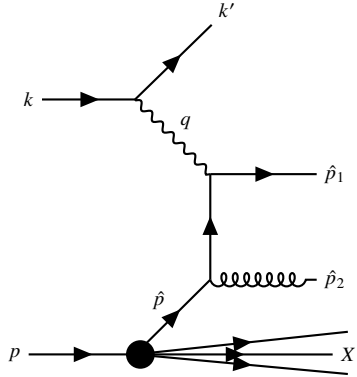

Deep inelastic scattering (DIS) processes enable the exploration of the structure of the proton and the scale evolution of quantum chromodynamics (QCD). Monte Carlo (MC) event generators can be used to model these processes from the hard partonic scattering all the way to the stable hadrons that can be observed in particle detectors.
General-purpose MC event generators employ parton showers to describe subsequent emissions, which transitions events from the high-energy core collision process to the lower-energy hadronization phase. 
Multi-jet merging is an algorithmic way of combining parton showers with multi-parton final states calculated from QCD scattering matrix elements (ME). As parton showers are also capable of producing such high-multiplicity states, a challenge in this procedure is to avoid double-counting phase-space points. \pythia \cite{Bierlich:2022pfr} implements LO and NLO merging based on the CKKW-L approach \cite{Lonnblad:2011xx, Lonnblad:2012ix}. 
Until now there have been no merging capabilities for DIS processes in \pythia. With the so called ``Power-Shower'' option \cite{Plehn:2005cq}, the default shower option is capable of producing jets and dijet events, such as one illustrated in Figure \ref{fig001:DISFeynman}. This approach is consistent with data in high-virtuality events, but struggles to match experimental results in lower virtuality events. This motivates the use of merging algorithms. 


\section{Multi-jet merging}

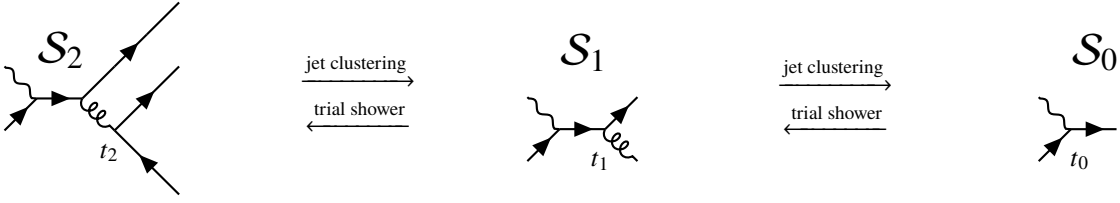
\begin{figure}[t]
 \centering
 \begin{minipage}[c]{0.4\textwidth}
 \begin{tikzpicture}[thick,scale=0.3, every node/.style={transform shape}]
        \begin{feynman}[large]
            \vertex (a);
            \vertex [below right=of a] (c);
            \vertex [below left=of c] (b);
            \vertex [right=of c] (d);
            \vertex [above right=of d] (e);
            \vertex [above right=of e] (f);
            \vertex [above right=of f] (g);
            \vertex [below right=of d] (h);
            \vertex [above right=of h] (k);
            \vertex [below right=of k] (m);
            \vertex [above right=of k] (l);
            \vertex [below right=of h] (i);
            \vertex [below right=of i] (j);

            \vertex [below=of d] (Q01);
            \vertex [below=of Q01] (Q00);
            \vertex [above=of d] (Q02);
            \vertex [above=of Q02] (Q03);
            \vertex [above=of e] (Q13);
            \vertex [below=of h] (Q10);   
            \vertex [below=of i] (Q20);   
            \vertex [above=of f] (Q23);   

            \node [scale=5.0] at (2.5, 0.75) {\( \mathcal{S}_2 \)};
            \node [scale=3.0] at (4.6,-3.8) {\( t_2 \)};

            \node [scale=3.0] at (15.5, -0.2) {\( \xrightarrow{\text{jet clustering}} \)};
            \node [scale=3.0] at (15.5, -2.2) {\( \xleftarrow{\text{trial shower}} \)};
            
            \diagram* {
                (a) -- [boson] (c), 
                (b) -- [fermion] (c) -- [fermion] (d),
                (d) -- [plain] (e) -- [fermion] (f) -- [plain] (g), 
                (d) -- [gluon] (h),
                (j) -- [fermion] (h),
                (h) -- [fermion] (l),
            };
        \end{feynman}
    \end{tikzpicture}

 \end{minipage}\hfill
 \begin{minipage}[c]{0.39\textwidth} 
  \begin{tikzpicture}[thick,scale=0.3, every node/.style={transform shape}]
        \begin{feynman}[large]
            \node[vertex] (a) at (0,0); 
            \vertex [below right=of a] (c);
            \vertex [below left=of c] (b);
            \vertex [right=of c] (d);
            \vertex [above right=of d] (e);
            \vertex [above right=of e] (f);
            \vertex [above right=of f] (g);
            \vertex [below right=of d] (h);
            \vertex [above right=of h] (k);
            \vertex [below right=of k] (m);
            \vertex [above right=of k] (l);
            \vertex [below right=of h] (i);
            \vertex [below right=of i] (j);

            \vertex [below=of d] (Q01);
            \vertex [below=of Q01] (Q00);
            \vertex [above=of d] (Q02);
            \vertex [above=of Q02] (Q03);
            \vertex [above=of e] (Q13);
            \vertex [below=of h] (Q10);   
            \vertex [below=of i] (Q20);   
            \vertex [above=of f] (Q23);   

            \node [scale=5.0] at (2.5, 2.0) {\(  \mathcal{S}_1 \)};
            \node [scale=3.0] at (3.2, -2.8) {\( t_1 \)};
            \node [scale=3.0] at (13.5, 1.0) {\( \xrightarrow{\text{jet clustering}} \)};
            \node [scale=3.0] at (13.5, -1.0) {\( \xleftarrow{\text{trial shower}} \)};
            
            \diagram* {
                (a) -- [boson] (c), 
                (b) -- [fermion] (c) -- [fermion] (d),
                (d) -- [fermion] (e), 
                (d) -- [gluon] (h),
            };
        \end{feynman}
    \end{tikzpicture}
  \end{minipage}\hfill
  \begin{minipage}[c]{0.1\textwidth}
   \begin{tikzpicture}[thick,scale=0.3, every node/.style={transform shape}]
    \begin{feynman}[large]
        \node[vertex] (a) at (0,0); 
        \vertex [below right=of a] (c);
        \vertex [below left=of c] (b);
        \vertex [right=of c] (d);
        \vertex [above right=of d] (e);
        \vertex [above right=of e] (f);
        \vertex [above right=of f] (g);
        \vertex [below right=of d] (h);
        \vertex [above right=of h] (k);
        \vertex [below right=of k] (m);
        \vertex [above right=of k] (l);
        \vertex [below right=of h] (i);
        \vertex [below right=of i] (j);

        \vertex [below=of d] (Q01);
        \vertex [below=of Q01] (Q00);
        \vertex [above=of d] (Q02);
        \vertex [above=of Q02] (Q03);
        \vertex [above=of e] (Q13);
        \vertex [below=of h] (Q10);   
        \vertex [below=of i] (Q20);   
        \vertex [above=of f] (Q23);   

        \node [scale=5.0] at (2.5, 2.0) {\( \mathcal{S}_0 \)};
        \node [scale=3.0] at (1.8, -2.8) {\( t_0 \)};
        
        \diagram* {
            (a) -- [boson] (c), 
            (b) -- [fermion] (c),
            (c) -- [fermion] (d), 
        };
    \end{feynman}
\end{tikzpicture}

  \end{minipage}%
 \caption{Illustration of how a merging algorithm reconstructs shower histories and produces shower states. }
 \label{fig002:merging_illustration}
\end{figure}

The basic idea of merging is to combine parton showers with parton-level events containing additional partons calculated in perturbative QCD, such that each will populate the part of phase space where they are accurate. 
Events calculated from QCD MEs can describe hard, well-separated jets, while parton showers generate collinear or soft, low-$p_\mathrm{T}$ emissions at good precision. The combination of these approaches is achieved by separating the phase space at some energy scale $t_\mathrm{MS}$, which defines soft and hard jets at a relevant scale, such as the jet transverse momentum. The merging scale $t_\mathrm{MS}$ also aims to assign the generation of hard jets to the ME events and soft partons to parton showers, with the definitions
\begin{equation*}
    \text{Hard } \text{jets}: p_\mathrm{T} > t_{\mathrm{MS}}. \qquad
    \text{Soft } \text{jets}: p_\mathrm{T} \leq t_{\mathrm{MS}}. 
\end{equation*}
In our setup, the value of $t_\mathrm{MS}$ is compared to the square root of the parton shower evolution variable, which we will call $t$. With the scale defined, a generic $N$-jet merging algorithm proceeds as follows: 
\begin{enumerate}
    \item Start from an event from a fixed-order ME sample. Enforce the merging scale cut, so that each additional final state parton obeys $t > t_\mathrm{MS}$. 
    \item Using a jet clustering algorithm, invert the shower kinematics to construct shower histories, obtain states $\{\mathcal{S}_0, \mathcal{S}_1 \dots \mathcal{S}_n \}$ and corresponding scales $\{t_0, t_1 \dots t_n \}$. Choose the most probable history. 
    \item Starting from state $\mathcal{S}_0$, perform trial emissions. Reject the event if the generated trial scale $t > t_i$. Reweight the event with PDF ratios and $\alpha_\mathrm{S}$ ratios. Accept the event and continue parton shower cascade from $t_n$ 
    \begin{itemize}
        \item if $n<N$, the jet multiplicity is not the highest, and the first emission is below $t_\mathrm{MS}$, or
        \item if $n=N$, the jet multiplicity is the highest.
    \end{itemize}
    Otherwise reject the whole event. 
\end{enumerate}
These steps are visualized in Figure \ref{fig002:merging_illustration}. 
The double-counting is removed by imposing trial showers between nodes corresponding to exact shower histories, and rejecting emissions above previously obtained scales, or above $t_\mathrm{MS}$. 
Another step in the algorithm is reweighting with the merging weight
\begin{equation}
    w_i = \Pi_{\mathcal{S}_i}(t_i, t_{i+1}) \frac{f(x_i,t_i)}{f(x_i,t_{i+1})} \frac{\alpha_\mathrm{S}(t_i)}{\alpha_\mathrm{S}(\mu_\mathrm{R})} , 
\end{equation}
which consists of ratios of parton distribution functions (PDFs) and strong coupling constants $\alpha_\mathrm{S}$. These factors make sure that the events get the same weight as they would in the parton shower approximation. This treatment includes higher-order corrections to soft-gluon emissions, normalizes events to the same cross section and correctly accounts for running of $\alpha_\mathrm{S}$. 

Two merging algorithms are employed: CKKW-L, which is the primary algorithm used in \pythia, and UMEPS \cite{Lonnblad:2012ng}, an improved version capable of retaining the leading-order cross section. In order to allow high-multiplicity parton-level events to contribute at low virtualities in the CKKW-L merging algorithm, a dynamic merging scale is defined similarly as in Ref.~\cite{Carli:2010cg}, 
\begin{equation}
    \label{eq01:DynamicTMS}
    t_\mathrm{MS} = \frac{\Tilde{t}_\mathrm{MS}}{\sqrt{1+ \frac{\Tilde{t}_\mathrm{MS}^2}{Q^2 S^2}}},
\end{equation}
with the parameters $\Tilde{t}_\mathrm{MS} = 5$ GeV to dictate the highest value of the merging scale and $S = 0.7$ to reduce the scale at low virtuality. A fixed merging scale is used with the UMEPS algorithm. The computational setup consists of these merging algorithms and scale choices, with parton-level events generated by \sherpa\footnote{https://gitlab.com/hpcgen/me} \cite{Gleisberg:2008ta,Hoche:2019flt, Bothmann:2023ozs}, the PDF set \texttt{NNPDF40\_lo\_pch\_as\_01180} \cite{NNPDF:2021njg}, and the \vincia sector-antenna shower \cite{Brooks:2020upa} in conjunction with its dedicated multi-jet merging framework \cite{Brooks:2020mab}. 

\section{Results}

We have simulated DIS events with up to $4$ additional jets with \sherpa. The results of multi-jet merging are compared to data from HERA experiments. We have produced two new \rivet jet analyses with data collected from experiments conducted in 1996 and 1997 with beam energies of $E_p = 820$ GeV and $E_e = 27.5$ GeV \cite{H1:2002qhb}, and from 2005 to 2007 with higher energies of $E_p = 920$ GeV and $E_p = 27.6$ GeV \cite{H1:2016goa}, both released by the H1 collaboration. 

\begin{figure}[h]
    \centering
    \begin{minipage}{.45\textwidth}
    \includegraphics[scale=0.6]{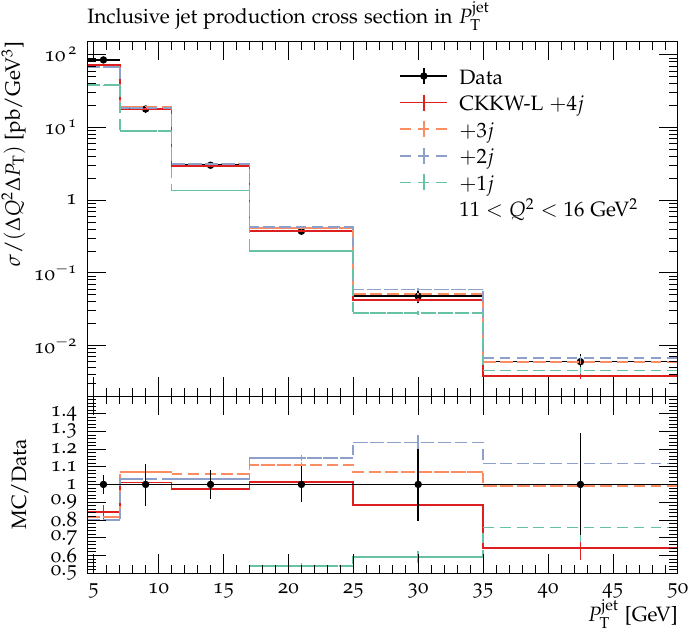}
    \end{minipage}
    \begin{minipage}{.45\textwidth}
    \includegraphics[scale=0.6]{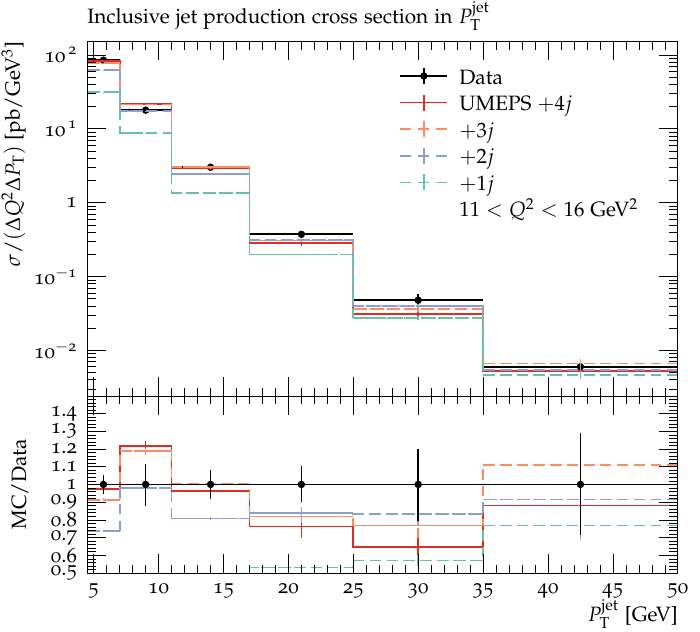}
    \end{minipage}
    \caption{Jet $p_\mathrm{T}$ distribution comparisons to H1 data for inclusive jet production cross sections with CKKW-L and UMEPS merging, from $+1$-jet merging up to $+4$-jet merging. }
    \label{fig006:Incjets4_CKKWL_UMEPS} 
\end{figure}

After the events with different number of outgoing partons are combined using the merging procedure described above, the parton showers are generated and partonic events hadronized providing a hadronic final state from which jets are reconstructed according to the $k_\mathrm{T}$-jet finder algorithm implemented in FastJet \cite{Cacciari:2011ma}. 
The identified jets typically originate from the high-$p_{\mathrm{T}}$ partons in the generated parton-level process. 
Figure \ref{fig006:Incjets4_CKKWL_UMEPS} shows jet production cross section comparisons from both merging algorithms, with a varying number or additional partons included in the fixed-order samples. Using $+1$-jet merging one cannot describe the data, since the simulation produces much lower jet cross sections. For inclusive distributions, $+2$-jet merging is already sufficient to describe the data within uncertainties. Adding further samples to the merging leads to approximately converging distributions. This reflects the perception that most of the inclusive jet cross-section contribution is obtained with parton-level events containing up to 3 partons. 

\begin{figure}
    \centering
    \begin{minipage}{.45\textwidth}
    \includegraphics[scale=0.6]{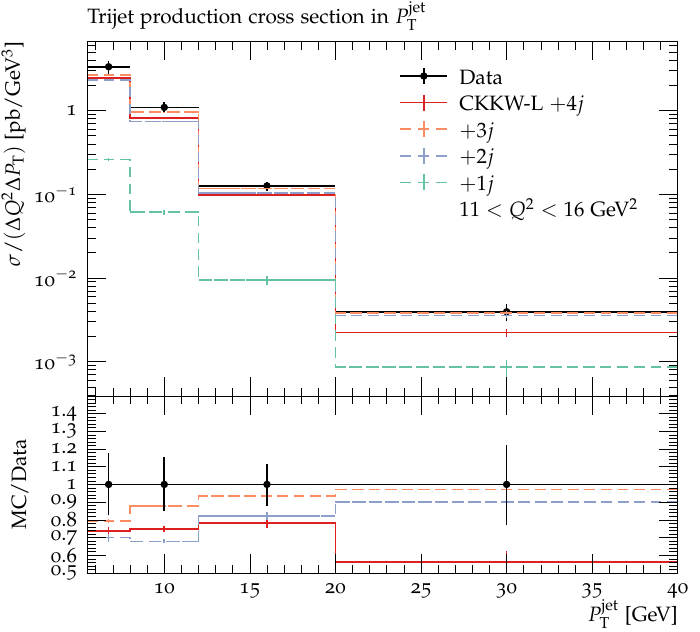}
    \end{minipage}
    \begin{minipage}{.45\textwidth}
    \includegraphics[scale=0.6]{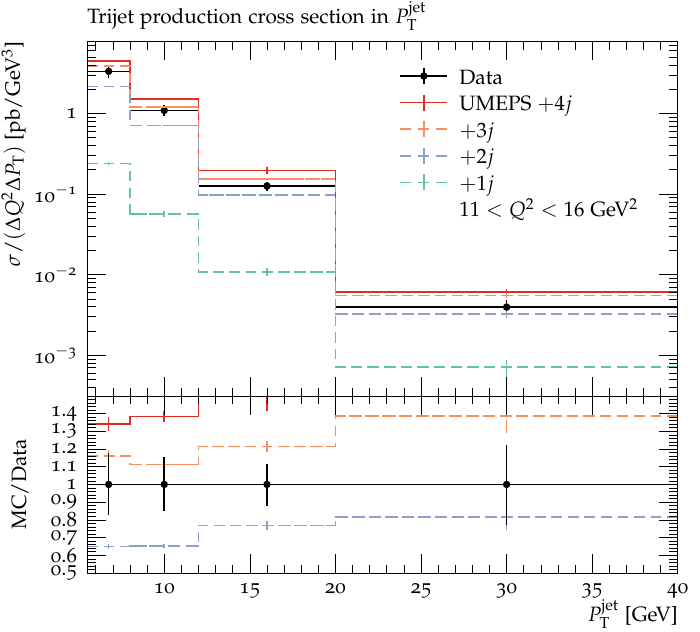}
    \end{minipage}
    \caption{Comparisons to H1 data for trijet $p_\mathrm{T}$ distributions with CKKW-L and UMEPS $+n$-jet predictions. }
    \label{fig008:Trijets4_CKKWL_UMEPS}
\end{figure}

Trijet distributions in Figure \ref{fig008:Trijets4_CKKWL_UMEPS} are obtained by calculating the cross sections in terms of the average $p_\mathrm{T}$ of the three hardest jets. For this higher jet multiplicity, even higher-multiplicity parton level samples are required. Compared to the inclusive jet distributions, there's still a considerable improvement in trijet cross sections when going from $+2$-jet to $+3$-jet merging. 

\newpage

\begin{figure}
    \centering
    \begin{minipage}{0.45\textwidth}
        \includegraphics[scale=0.6]{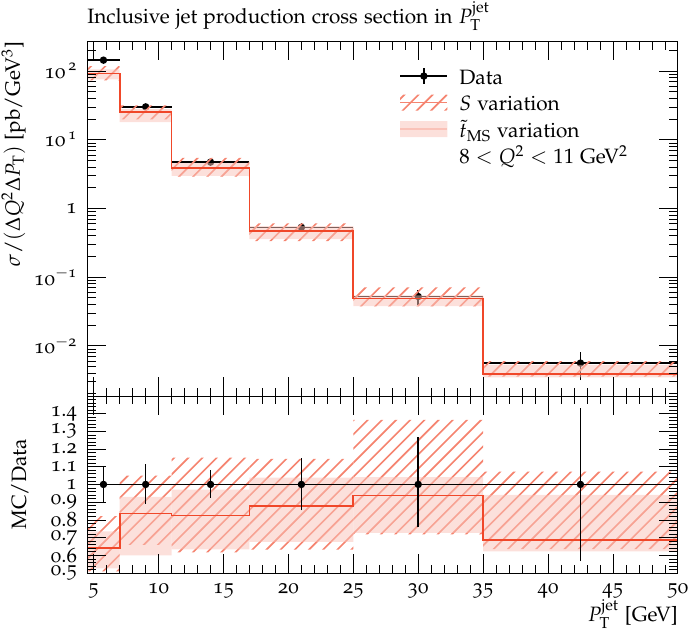}   
    \end{minipage}
    \begin{minipage}{0.45\textwidth} 
        \includegraphics[scale=0.6]{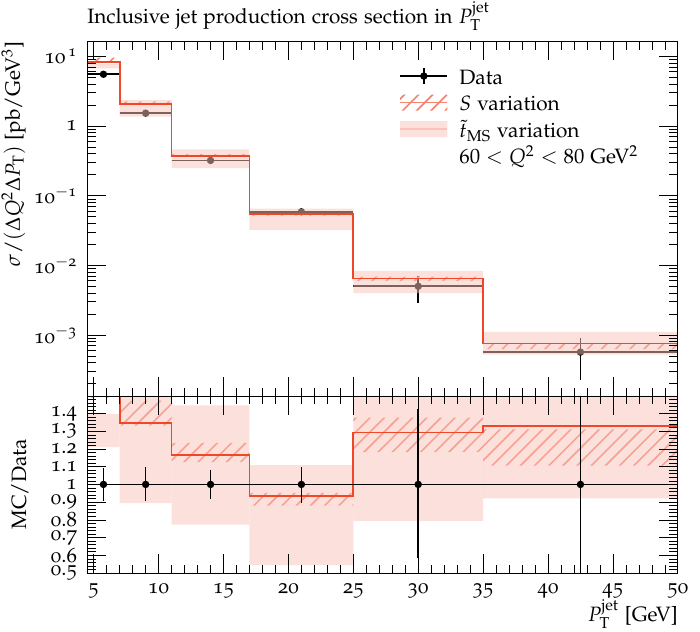}   
    \end{minipage}
    \caption{Inclusive jet cross section differential in jet $p_\mathrm{T}$. Variations of parameters $S = 0.7 \pm 0.1$ and $\tilde{t}_\mathrm{MS} = 5^{+4}_{-2}$ GeV are shown as error bands. Left: low-$Q^2$ events. Right: high-$Q^2$ events.}
    \label{fig005:DynamicTMSvariations}
\end{figure}

In Figure \ref{fig005:DynamicTMSvariations} we present the cross sections resulting from $t_\mathrm{MS}$ and $S$ variations. Dynamic merging scale has a larger impact at low virtuality and indeed the considered variation of the parameter $S$ leads to a larger uncertainty. These variations are an essential part of uncertainty analyses of simulations as jet properties are sensitive to changes in these parameters. 

\section{Conclusions}

We present the implementation of multi-jet merging in DIS with \pythia, using the \vincia parton shower. The CKKW-L and UMEPS algorithms are employed with parton-level events from \sherpa. Results are compared to HERA data, and we find good agreement of differential multi-jet cross sections across a wide range of photon virtuality when including at least 3 final-state partons for the inclusive jet cross sections. 

\newpage

\acknowledgments
This research was funded through the Research Council of Finland project No. 330448 and No. 331545, as a part of the Center of Excellence in Quark Matter of the Research Council of Finland. \\ We acknowledge grants of computer capacity from the Finnish Grid and Cloud Infrastructure (persistent identifier urn:nbn:fi:research-infras-2016072533). \\ Part of the computations were carried out on the PLEIADES computing cluster at the University of Wuppertal, supported by the Forschungsgemeinschaft (DFG, grant No. INST 218/78-1 FUGG) and the Bundesministerium für Bildung und Forschung (BMBF).

\end{document}